# Use of Combined Hartree-Fock-Roothaan Theory in Evaluation of Lowest States of K [Ar]4s⁰3d¹ and Cr⁺ [Ar]4s⁰3d⁵ Isoelectronic Series Over Noninteger n-Slater Type Orbitals

I.I. Guseinov, M. Ertürk and E. Şahin

*Department of Physics, Faculty of Arts and Sciences, Onsekiz Mart University, Çanakkale, Turkey*

**Abstract**

By the use of integer and noninteger n-Slater Type Orbitals in combined Hartree-Fock-Roothaan method, self consistent field calculations of orbital and lowest states energies have been performed for the isoelectronic series of open shell systems K $[Ar]4s^03d^1$ $(^2D)$ (Z=19-30) and $Cr^+$ $[Ar]4s^03d^5$ $(^6S)$ (Z=24-30). The results of calculations for the orbital and total energies obtained from the use of minimal basis sets of integer- and noninteger n-Slater Type Orbitals are given in tables. The results are compared with the extended-basis Hartree-Fock computations. The orbital and total energies are in good agreement with those presented in the literature. The results are accurately and considerably can be useful in the application of non-relativistic and relativistic combined Hartree-Fock-Roothaan approach for heavy atomic systems.

**Key Words:** Hartree-Fock-Roothaan equations, Noninteger n-Slater type orbitals, Open shell theory, Isoelectronic series

## 1. Introduction

The Hartree-Fock-Roothaan (HFR) or basis-set expansion method is a convenient and powerful tool for the study of electronic structure of atoms and molecules [1-3]. It is well known that the open shell HFR theory [4] and its extensions by others [5-15] are not applicable to any state of a single configuration, which has any symmetry of open shells. Most of these methods apply only to particular couplings conditions. In a previous paper [16], one of the authors has suggested the combined HFR (CHFR) method for arbitrary open shell systems. In Refs. [17, 18], the CHFR method numerically has been confirmed for He and some first row atoms in single and double zeta approximations. The choice of the basis functions is very important when the CHFR method is employed, because it determines both the computational efficiency and accuracy of the results obtainable within a given approaches. The most frequently used basis functions for atomic calculations are Slater type orbitals (STOs) [19] defined as

$$\chi_{n^*lm}(\zeta,\vec{r}) = \frac{(2\zeta)^{n^*+\frac{1}{2}}}{\left[\Gamma(2n^*+1)\right]^{\frac{1}{2}}} r^{n^*-1} e^{-\zeta r} S_{lm}(\theta,\varphi). \tag{1}$$

Here, $\Gamma(x)$ and $S_{lm}(\theta,\varphi)$ are the gamma function and complex or real spherical harmonics, respectively; $\zeta > 0$ is the orbital exponent. The quantity $n^*$ occurring in Eq. (1) is positive integer or noninteger principal quantum number of STOs (ISTOs and NSTOs). It should be noted that the STOs allow an adequate representation of important properties of the electronic wave function, such as the cusps at the nuclei [20] and the exponential decay at large distances [21]. The determination of nonlinear parameters $n^*$ and $\zeta$ appearing in Eq. (1) is very important for describing the atomic orbitals.

Over the years, much progress has been made in using ISTOs and NSTOs as atomic basis functions in the electronic structure calculations of atoms [22-34]. The ground state energies of isoelectronic series with the open shells for first-row transition atoms were calculated by Clementi-Roetti [35] and C. F. Fischer [36] using Hartree-Fock limit (HFL) in the ISTOs basis sets and numerical Hartree-Fock (NHF) approaches, respectively. As we know from the literature that the isoelectronic series of atoms have not been investigated exhaustively using the NSTOs basis sets, particularly for the heavy atoms and ions.

The purpose of the present paper is to perform, as an application of the open shell CHFR method, the calculations for the lowest states of the isoelectronic series of K [Ar]$4s^0 3d^1$ and Cr$^+$ [Ar]$4s^0 3d^5$ configurations using ISTOs and NSTOs in the minimal basis set approximation. The computational method is described in the next section. Hartree atomic units used throughout this work.

## 2. Theory and Computational Method

It is well known that the total energy functional in HFR method is the average of the energies of all the important symmetry-adapted configuration state functions. The orbitals are optimized so that the energy functional reaches its minimum. According to CHFR theory the postulated energy for arbitrary number of closed and open shell electronic configurations, in the case of atoms is given by [16]:

$$E(LS) = 2\sum_i^n f_i h_i + \sum_{i,j,k,l}^n \left(2 A_{kl}^{ij} J_{kl}^{ij} - B_{kl}^{ij} K_{kl}^{ij}\right). \tag{2}$$

Here, $f_i$ is the fractional occupancy of shell $i$. The $h_i$, $J_{kl}^{ij}$ and $K_{kl}^{ij}$ are the one electron, and Coulomb and exchange integrals, respectively. The coupling-projection (CP) coefficients $A_{kl}^{ij}$ and $B_{kl}^{ij}$ appearing in Eq. (2) in the case of closed-closed and closed-open shells interaction are determined as

$$A_{kl}^{ij} = B_{kl}^{ij} = f_i f_k \delta_{ij} \delta_{kl}. \tag{3}$$

The necessity for the determination of the open-open shells interaction CP coefficients arises of using the following relation of the expectation value for the total energy:

$$E(^{2S+1}L) = \frac{1}{(2L+1)(2S+1)} \sum_{M_L M_S} \int \Psi_{M_L M_S}^{LS*} \hat{H} \Psi_{M_L M_S}^{LS} d\tau. \tag{4}$$

In order to find the CP coefficients, we should compare the results obtained from Eq. (2) with those in Eq. (4) (See Refs. [16-18] and references therein for more details about these coefficients). The wave functions $\Psi_{M_L M_S}^{LS}$ occurring in Eq. (4) can be constructed by the use of determinantal Slater [37] or determinantal modified Slater method (see Sec. (4) of [16]). We notice that the CP coefficients have the following symmetry properties:

$$A_{kl}^{ij} = A_{ij}^{kl}, \quad A_{kl}^{ij} = A_{lk}^{ji}, \tag{5}$$

$$B_{kl}^{ij} = B_{ij}^{kl}, \quad B_{kl}^{ij} = B_{lk}^{ji}. \tag{6}$$

The CP coefficients, in the case of isoelectronic series K $[Ar]4s^03d^1$ $(^2D)$, which have a single electron in open shell, can be determined by the use of Eq. (3). The values of open-open shell CP coefficients for the lowest state of isoelectronic series $Cr^+$ are given in table 1. In table 1, the Slater atomic orbitals are denoted by

$nlm$: 100  200  211  210  21−1  300  311  310  31−1  322  321  320  32−1  32−2
$i$:    1    2    3    4     5    6    7    8     9    10   11   12   13    14

Now we rewrite Eq. (2) in terms of the linear combination coefficients C. Then, we obtain for the energy functional the following algebraic relation:

$$E = 2\sum_q (\rho h)_{qq} + \sum_{ij,kl} \sum_{pq} (2 C_{ip}^\dagger A_{kl}^{ij} \Gamma_{kl}^{pq} C_{qj} - C_{ip}^\dagger B_{kl}^{ij} \Lambda_{kl}^{pq} C_{qj}), \tag{7}$$

where $\rho = CfC^\dagger$ and

$$\Gamma_{kl}^{pq} = (C^\dagger I^{pq} C)_{kl} \tag{8}$$

$$\Lambda_{kl}^{pq} = (C^\dagger K^{pq} C)_{kl}. \tag{9}$$

Two electron integrals $I_{kl}^{pq}$ and $K_{kl}^{pq}$ occurring in these equations are defined as

$$I_{kl}^{pq} = \iint \chi_p^*(x_1) \chi_k^*(x_2) \frac{1}{r_{21}} \chi_q(x_1) \chi_l(x_2) dv_1 dv_2 \tag{10}$$

$$K_{kl}^{pq} = \iint \chi_p^*(x_1)\chi_k^*(x_2)\frac{1}{r_{21}}\chi_l(x_1)\chi_q(x_2)dv_1dv_2. \tag{11}$$

The Coulomb and exchange integrals defined by Eqs. (10) and (11), respectively, have the same form except the indices. These one-center two electron integrals over the NSTOs basis sets can be evaluated by the use of relations in terms of the gamma and hypergeometric functions presented in Refs. [31, 38].

Recently, we have constructed a self-consistent field program in Mathematica international mathematical software based on the CHFR method using ISTOs and NSTOs basis sets [39]. All the nonlinear parameters were optimized with the help of quasi-Newton method. Initial values of the nonlinear parameters occurring in NSTOs were taken from the ISTOs results. The values of nonlinear parameters obtained are used, then, in the optimization of NSTOs nonlinear parameters. We observed that the final virial ratios do not deviate from the exact value of -2 by more $10^{-7}$. All of the values of parameters tabulated have nine decimals. Because of this, the ratios obtained are so much more precise than those of Ref. [35]. We notice that our optimizations in all of the calculations are sufficiently accurate.

## 3. Results and Discussion

The CHFR method described above and developed to date appears to yield good results when applied in calculations for atomic systems. In case of the ISTOs and NSTOs basis sets, the total energies ($E_{ISTO}$ and $E_{NSTO}$), their differences ($\Delta E_{E_{ISTO}-E_{NSTO}}$) and energy differences $E_{NSTO}$ and HFL ($\Delta E_{E_{NSTO}-E_{HFL}}$) for the isoelectronic series K[Ar]$4s^03d^1$ ($^2D$) and Cr$^+$[Ar] $4s^03d^5$ ($^6S$) configurations are given in tables 2 and 3, respectively. We see from tables that the total energy differences $\Delta E_{E_{ISTO}-E_{NSTO}}$ increases as the atomic number Z increases. We also observed that the efficiency of NSTOs increases rapidly with the increased atomic number Z with respect to the ISTOs. The comparison of our results obtained in NSTOs basis sets with corresponding Clementi-Roetti's HFL energies [35] shows that the energy differences $\Delta E_{E_{NSTO}-E_{HFL}}$ are almost not changed.

The total energy errors for the ISTOs and NSTOs basis sets relative to the NHF [36] for the isoelectronic series K and Cr$^{+1}$ are shown in figures 1 and 2, respectively. We call attention to the fact that the total energy errors $\Delta E_{E_{ISTO}-E_{NHF}}$ for the ISTOs basis sets increase smoothly with an increase in both the atomic number Z and number of electrons. In the NSTOs basis sets calculations, on the contrary, the trend of total energy errors $\Delta E_{E_{NSTO}-E_{NHF}}$ is in opposite direction consistently. As seen from figures 1 and 2, the results of NSTOs basis sets calculations are about 3-4 times more accurate than those ISTOs basis set.

We notice that the Clementi-Roetti HFL energies [35] for the Z=28 in isoelectronic series of configuration K[Ar]$4s^0 3d^1$ $(^2D)$ is incorrect. It should be noted that for Z=20 in K isoelectronic series Clementi-Roetti configuration (K[Ar]$4s^1 3d^0$) is different from our configuration (K[Ar]$4s^0 3d^1$).

In tables 4 and 5, we present orbital energies obtained from the NSTOs basis sets for the isoelectronic series K and $Cr^{+1}$. For these isoelectronic series, all the orbital energies are improved in NSTOs calculations within the minimal basis framework.

Table 6 lists the optimum nonlinear parameters $n^*$ and $\zeta$ for the Z=25 in K and $Cr^{+1}$ isoelectronic series. It should be noted that the deviation from nominal values of principal quantum numbers is largest for 3d orbital, i.e., $1.93 \leq n^*_{3d} \leq 2.51$. The improvement of the NSTOs basis sets by the extension of the $n^*$ from integer to noninteger values is due to the fact that these optimal values are significantly different from the nominal values of orbitals.

The forcefulness of the NSTOs basis sets in calculations of the heavy atoms has been studied for isoelectronic series K and $Cr^{+1}$. We showed that the efficiency of the NSTOs basis sets is increased especially with increase the atomic number Z in isoelectronic series of K[Ar]$4s^0 3d^1$ $(^2D)$ and $Cr^+$[Ar] $4s^0 3d^5$ $(^6S)$ configurations. NSTOs total and orbital energies are closer to the corresponding NHF values than the ISTOs ones. The size of the present noninteger n-Slater orbitals is smaller than that of the usual extended integer n-Slater orbitals presented in literature. This reduction considerably extends the range of applications of the NSTOs in calculations of atomic properties. We notice that the optimization process in our calculations is not restricted by condition $n^* \geq 1$ for fulfilling the cusp value and asymptotic long-range behavior of orbitals. We plan further to take into account these criteria of atomic orbitals. Work is in progress in our group for the analytical CHFR calculations of ground and excited states of heavy atoms and their ions. It should be noted that the analytical solutions of CHFR equations, Eqs. (2) and (7), can be performed for the arbitrary states of electronic configurations which have any number of open shells (whatever their symmetry is). The CHFR method can be of considerable help and importance in the simplification of open shell atomic structure calculations. This method has, in our opinion, an advantage and simplex over standard approaches presented in literature. The CHFR method and a computer code are presented which allow more forceful computation of HFL energies. The resulting wave functions and parameters are available by request through e-mail addresses: ihuseyin@comu.edu.tr or merturk@comu.edu.tr.

**Acknowledgments**

This work was financially supported by the TÜBİTAK project No. TBAG-2396 (103T172). One of us, EŞ, thanks TUBITAK-BIDEB for financial support for his PhD. education.

Table 1. The values of open-open shells coupling-projection coefficients for isoelectronic series of the lowest state electronic configuration $Cr^+$ [Ar] $4s^03d^5$ $(^6S)$

| i  j  | K  l  | $A_{kl}^{ij} = A_{ij}^{kl}$ | $B_{kl}^{ij} = B_{ij}^{kl}$ |
|-------|-------|-----------------------------|-----------------------------|
| 10 10 | 11 11 | ¼ | ½ |
|       | 12 12 | ¼ | ½ |
|       | 13 13 | ¼ | ½ |
|       | 14 14 | ¼ | ½ |
| 11 11 | 12 12 | ¼ | ½ |
|       | 13 13 | ¼ | ½ |
|       | 14 14 | ¼ | ½ |
| 12 12 | 13 13 | ¼ | ½ |
|       | 14 14 | ¼ | ½ |
| 13 13 | 14 14 | ¼ | ½ |

Table 2. The ISTO ($E_{ISTO}$) and NSTO ($E_{NSTO}$) total energies, total energy differences $\Delta E_{E_{ISTO}-E_{NSTO}}$ and $\Delta E_{E_{NSTO}-E_{HFL}}$ for the isoelectronic series K [Ar]$4s^03d^1$ $(^2D)$

| Atomic Number (Z) | -$E_{ISTO}$ | -$E_{NSTO}$ | $\Delta E_{E_{ISTO}-E_{NSTO}}$ | $\Delta E_{E_{NSTO}-E_{HFL}}$ |
|---|---|---|---|---|
| 19 | 598.004932042 | 598.851381634 | 0.846449592 | - |
| 20 | 675.292908562 | 676.252721310 | 0.959812748 | 0.317168690 |
| 21 | 757.792879118 | 758.849276755 | 1.056397637 | 0.242913245 |
| 22 | 845.390053687 | 846.509362081 | 1.119308394 | 0.239847919 |
| 23 | 938.016084491 | 939.207319355 | 1.191234864 | 0.238080645 |
| 24 | 1035.659729808 | 1036.932938799 | 1.273208991 | 0.237261201 |
| 25 | 1138.314776362 | 1139.680087694 | 1.365311332 | 0.236412306 |
| 26 | 1245.977048758 | 1247.444559744 | 1.467510986 | 0.236240256 |
| 27 | 1358.643498206 | 1360.223277271 | 1.579779065 | 0.237722729 |
| 28 | 1476.311794713 | 1478.013899031 | 1.702104318 | - |
| 29 | 1598.980104698 | 1600.814595675 | 1.834490977 | 0.239904325 |
| 30 | 1726.646954721 | 1728.623908862 | 1.976954141 | 0.240791138 |

Table 3. The ISTO ($E_{ISTO}$) and NSTO ($E_{NSTO}$) total energies, total energy differences $\Delta E_{E_{ISTO}-E_{NSTO}}$ and $\Delta E_{E_{NSTO}-E_{HFL}}$ for the isoelectronic series $Cr^+$ [Ar] $4s^03d^5$ $(^6S)$

| Atomic Number (Z) | -$E_{ISTO}$ | -$E_{NSTO}$ | $\Delta E_{E_{ISTO}-E_{NSTO}}$ | $\Delta E_{E_{NSTO}-E_{HFL}}$ |
|---|---|---|---|---|
| 24 | 1040.506195491 | 1042.636094901 | 2.129899411 | 0.502705099 |
| 25 | 1146.418778218 | 1148.633507128 | 2.214728910 | 0.474792872 |
| 26 | 1257.856029647 | 1260.156391428 | 2.300361781 | 0.457708572 |
| 27 | 1374.789004246 | 1377.181383800 | 2.392379554 | 0.446916200 |
| 28 | 1497.202404334 | 1499.694945177 | 2.492540843 | 0.439554823 |
| 29 | 1625.086610845 | 1627.688157939 | 2.601547093 | 0.431942061 |
| 30 | 1758.434948335 | 1761.154680347 | 2.719732012 | 0.431219652 |

Table 4. The NSTO orbital energies $(\varepsilon_{NSTO})$ for the lowest states $(^2D)$ of isoelectronic series K from Z=19 to Z=30

| Z | 19 | 20 | 21 | 22 | 23 | 24 | 25 | 26 | 27 | 28 | 29 | 30 |
|---|---|---|---|---|---|---|---|---|---|---|---|---|
| $-\varepsilon_{1s}$ | 133.603651258 | 149.445801069 | 166.438177745 | 184.487104267 | 203.553605446 | 223.630444090 | 244.714652843 | 266.804425631 | 289.898648114 | 313.996500667 | 339.097385523 | 365.200830066 |
| $-\varepsilon_{2s}$ | 14.527120013 | 16.849342760 | 19.579452447 | 22.627510239 | 25.947895215 | 29.531782211 | 33.375355895 | 37.476297577 | 41.833107950 | 46.444697164 | 51.310255845 | 56.429144725 |
| $-\varepsilon_{2p}$ | 11.546200130 | 13.650400657 | 16.161008648 | 18.987356287 | 22.084638327 | 25.444311400 | 29.062752228 | 32.937780295 | 37.068012490 | 41.452453486 | 46.090369027 | 50.981182656 |
| $-\varepsilon_{3s}$ | 1.827697709 | 2.312210347 | 3.093788374 | 4.047669496 | 5.129549919 | 6.331782843 | 7.651310971 | 9.086435448 | 10.636113250 | 12.299627093 | 14.076463021 | 15.966229441 |
| $-\varepsilon_{3p}$ | 1.035139568 | 1.424593499 | 2.106449792 | 2.954319846 | 3.928433380 | 5.022108584 | 6.232674030 | 7.558616900 | 8.998997265 | 10.553159136 | 12.220621649 | 14.001013606 |
| $-\varepsilon_{3d}$ | 0.056565108 | 0.326254285 | 0.875393640 | 1.565447443 | 2.381668575 | 3.319089997 | 4.375106731 | 5.548039832 | 6.836733031 | 8.240340802 | 9.758222895 | 11.389865889 |

Table 5. The NSTO orbital energies $(\varepsilon_{NSTO})$ for the lowest states $(^6S)$ of isoelectronic series $Cr^{+1}$ from Z=24 to Z=30

| Z | 24 | 25 | 26 | 27 | 28 | 29 | 30 |
|---|---|---|---|---|---|---|---|
| $-\varepsilon_{1s}$ | 220.595019376 | 241.141101900 | 262.723648049 | 285.326617591 | 308.943013703 | 333.569080966 | 359.202552327 |
| $-\varepsilon_{2s}$ | 26.375409912 | 29.667863580 | 33.258682357 | 37.128132456 | 41.267218875 | 45.670924483 | 50.336094178 |
| $-\varepsilon_{2p}$ | 22.301117799 | 25.369558579 | 28.734013305 | 32.375242950 | 36.284563846 | 40.457184582 | 44.890126812 |
| $-\varepsilon_{3s}$ | 3.444607661 | 4.371169138 | 5.450280137 | 6.663125942 | 8.001499573 | 9.461049106 | 11.039175810 |
| $-\varepsilon_{3p}$ | 2.216167820 | 3.041733119 | 4.013835251 | 5.116176253 | 6.341803787 | 7.687074998 | 9.149830560 |
| $-\varepsilon_{3d}$ | 0.529619234 | 1.187798200 | 1.984292134 | 2.908309596 | 3.954801558 | 5.120898920 | 6.404779781 |

Table 6. The nonlinear parameters of the NSTOs basis sets for the atomic number Z=25 in isoelectronic series of K and $Cr^+$

| Atomic Number Z=25 | K [Ar]$4s^03d^1$ $(^2D)$ | | $Cr^+$ [Ar] $4s^03d^5$ $(^6S)$ | |
|---|---|---|---|---|
| | $n^*$ | $\zeta$ | $n^*$ | $\zeta$ |
| $n_{1s}$ | 0.996452092 | 24.494169683 | 0.996475180 | 24.496025348 |
| $n_{2s}$ | 2.585633206 | 11.488829069 | 2.579379023 | 11.441813217 |
| $n_{2p}$ | 1.877837774 | 9.812004191 | 1.876183273 | 9.783910946 |
| $n_{3s}$ | 3.202373740 | 4.783804983 | 2.801893062 | 4.065240858 |
| $n_{3p}$ | 2.882563193 | 4.204804891 | 2.502967207 | 3.460314356 |
| $n_{3d}$ | 2.340092169 | 3.358124840 | 2.042644122 | 2.421278904 |

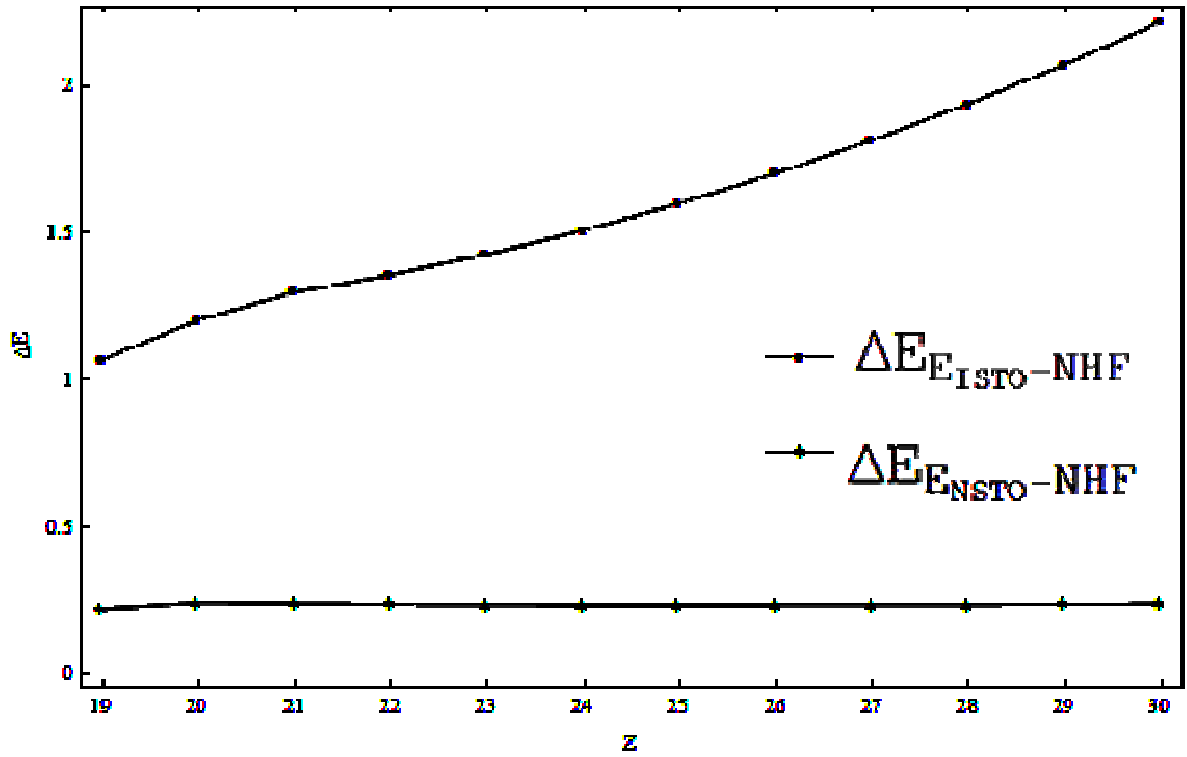

Figure 1. The energy differences $\Delta E_{E_{ISTO-NHF}}$ and $\Delta E_{E_{NSTO-NHF}}$ as a function of atomic number Z for the lowest states $(^2D)$ of isoelectronic series K (the results of NHF are obtained from Ref. [36])

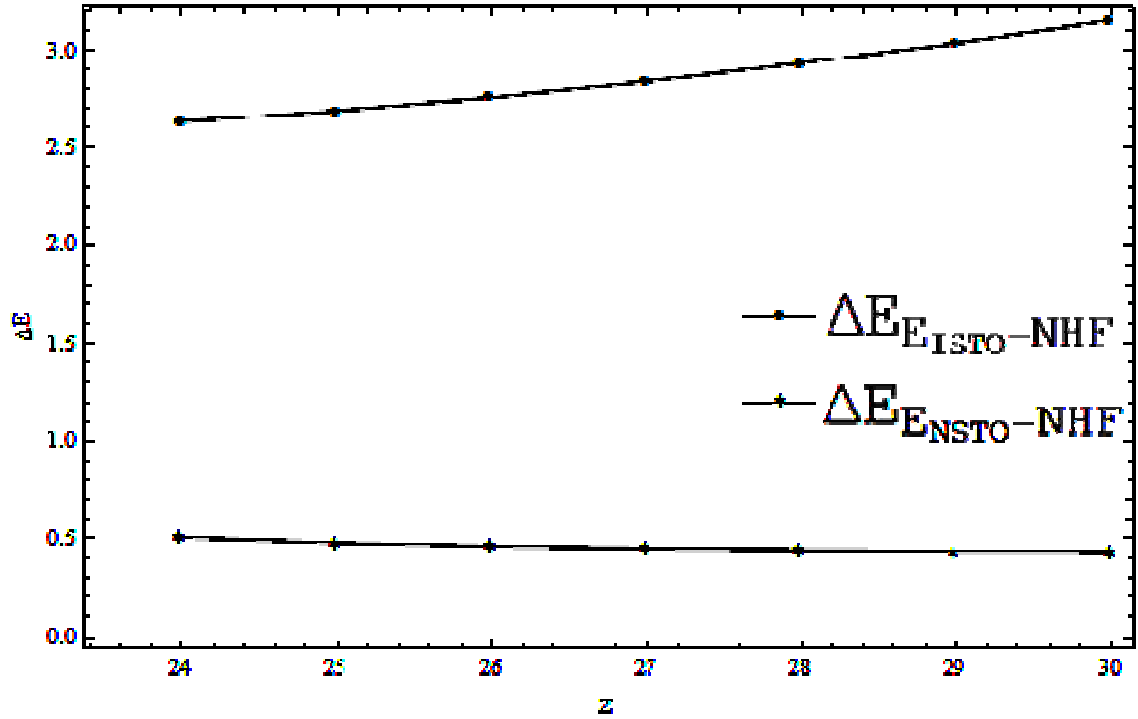

Figure 2. The energy differences $\Delta E_{E_{ISTO-NHF}}$ and $\Delta E_{E_{NSTO-NHF}}$ as a function of atomic number Z for the lowest states $\left(^6S\right)$ of isoelectronic series $Cr^{+1}$ (the results of NHF are obtained from Ref. [36])